# 4MOST Scientific Operations

C. Jakob Walcher[1]
Manda Banerji[2]
Chiara Battistini[3]
Cameron P. M. Bell[1]
Olga Bellido-Tirado[1]
Thomas Bensby[4]
Joachim M. Bestenlehner[5]
Thomas Boller[6]
Joar Brynnel[1]
Andrew Casey[7]
Cristina Chiappini[1]
Norbert Christlieb[3]
Ross Church[4]
Maria-Rosa L. Cioni[1]
Scott Croom[8]
Johan Comparat[6]
Luke J. M. Davies[9]
Roelof S. de Jong[1]
Tom Dwelly[6]
Harry Enke[1]
Sofia Feltzing[4]
Diane Feuillet[10]
Morgan Fouesneau[10]
Dominic Ford[4]
Steffen Frey[1]
Eduardo Gonzalez-Solares[2]
Alain Gueguen[6]
Louise Howes[4]
Mike Irwin[2]
Jochen Klar[1]
Georges Kordopatis[11]
Andreas Korn[12]
Mirko Krumpe[1]
Iryna Kushniruk[4]
Man I Lam[1]
James Lewis[2]
Karin Lind[12]
Jochen Liske[13]
Jon Loveday[14]
Vincenzo Mainieri[15]
Sarah Martell[16]
Gal Matijevic[1]
Richard McMahon[2]
Andrea Merloni[6]
David Murphy[2]
Florian Niederhofer[1]
Peder Norberg[17]
Alexander Pramskiy[3]
Martino Romaniello[15]
Aaron S. G. Robotham[9]
Florian Rothmaier[3]
Gregory Ruchti[4]
Olivier Schnurr[1, 18]
Axel Schwope[1]
Scott Smedley[19]
Jenny Sorce[20, 1]
Else Starkenburg[1]
Ingo Stilz[3]
Jesper Storm[1]
Elmo Tempel[21, 1]
Wing-Fai Thi[6]
Gregor Traven[4]
Marica Valentini[1]
Mario van den Ancker[15]
Nicholas Walton[2]
Roland Winkler[1]
C. Clare Worley[2]
Gabor Worseck[22]

[1] Leibniz-Institut für Astrophysik Potsdam (AIP), Germany
[2] Institute of Astronomy, University of Cambridge, UK
[3] Zentrum für Astronomie der Universität Heidelberg/Landessternwarte, Germany
[4] Lund Observatory, Lund University, Sweden
[5] Physics and Astronomy, University of Sheffield, UK
[6] Max-Planck-Institut für extraterrestrische Physik, Garching, Germany
[7] School of Physics and Astronomy, Monash University, Melbourne, Australia
[8] Sydney Institute for Astronomy, University of Sydney, Australia
[9] International Centre for Radio Astronomy Research / University of Western Australia, Perth, Australia
[10] Max-Planck-Institut für Astronomie, Heidelberg, Germany
[11] Observatoire de la Côte d'Azur, Nice, France
[12] Department of Physics and Astronomy, Uppsala universitet, Sweden
[13] Hamburger Sternwarte, Universität Hamburg, Germany
[14] University of Sussex, Brighton, UK
[15] ESO
[16] School of Physics, University of New South Wales, Sydney, Australia
[17] Department of Physics, Durham University, UK
[18] Cherenkov Telescope Array Observatory, Bologna, Italy
[19] Australian Astronomical Optics – Macquarie, Sydney, Australia
[20] Centre de Recherche Astrophysique de Lyon, France
[21] Tartu Observatory, University of Tartu, Estonia
[22] Institut für Physik und Astronomie, Universität Potsdam, Germany

The 4MOST instrument is a multi-object spectrograph that will address Galactic and extragalactic science cases simultaneously by observing targets from a large number of different surveys within each science exposure. This parallel mode of operation and the survey nature of 4MOST require some distinct 4MOST-specific operational features within the overall operations model of ESO. The main feature is that the 4MOST Consortium will deliver, not only the instrument, but also contractual services to the user community, which is why 4MOST is also described as a facility. This white paper concentrates on information particularly useful to answering the forthcoming Call for Letters of Intent.

## Operational context and requirements

4MOST is conceived as a survey facility that comprises the instrument and associated operations services. The largest fraction of the observing time on 4MOST will be allocated within a unique operational concept in which five-year Public Surveys from both the Consortium and the ESO community will be combined and observed in parallel during each exposure. These Surveys are jointly called Participating Surveys. ESO community members can also choose not to participate in this joint observing strategy by proposing a Non-Participating Survey. More details about the definitions and the selection procedures for Participating and Non-Participating Surveys can be found in the overview paper by de Jong et al., p. 3.

In the parallel observing mode, 4MOST will obtain spectra to serve many different science cases simultaneously. Parallel observing thus enables efficient use of 4MOST for surveys that have complementary observing conditions requirements and/or a target density lower than the 4MOST multiplexing capability. It also implies that surveys have to agree on a common survey strategy and prepare Observation Blocks (OBs) jointly. As a consequence, Participating Surveys will not explicitly choose the atmospheric conditions under which they wish to observe their targets. Rather, the design of the common survey strategy will be driven by observational success criteria.

These could, for example, be requirements on the signal-to-noise ratio (S/N) per target, or on the sky area to be covered. An additional consideration is that, owing to the nature of multi-object spectrographs, the spectra of targets will partially overlap on the detector (crosstalk between neighbouring fibres on the CCD). This implies that all Participating Surveys will have to fully share the raw data as well as the calibrated spectra in order to be able to assess and mitigate the impact of this cross-talk effect on their science.

4MOST operations have been designed to work within ESO's La Silla Paranal Observatory framework, with as few changes to infrastructure and processes as possible. Still, two main differences from the standard ESO science operations model are necessary: (1) a joint science team for all Participating Surveys (i.e., including Community Surveys as well as those from the instrument-building Consortium); (2) common centralised tasks in observation preparation and data management that are provided as a service by the Consortium.

The survey nature of 4MOST operations means that Targets of Opportunity or time-constrained observations on timescales shorter than a few days cannot be accommodated. However, transients that are numerous enough that they fall in randomly distributed 4MOST pointings can be observed if they can be included into the data stream with a few days lead time. Also, targets that require re-visits with a certain cadence can, in principle, be accommodated, in particular for deep fields requiring many visits. Towards the end of this paper we discuss the distinct case of Non-Participating Surveys, i.e., surveys that wish to use 4MOST in single survey mode.

Within 4MOST we define three data levels as follows: Level zero (L0) data are raw data, calibration data, environmental data, and log files; Level one (L1) data are one-dimensional (1D), calibrated, science-ready spectra extracted from the raw data; and Level two (L2) data are products resulting from the science analysis of 1D spectra, in particular physical properties of 4MOST targets. Examples of L2 data include elemental abundances for stars or redshifts and emission line fluxes for galaxies. L2 products also include spectra stacked over several OBs. L2 products that are to be delivered to ESO in Phase 3 are deliverable L2 (DL2) products. Any survey may also generate additional L2 (AL2) products.

## Organisational setup and roles

An organisation chart for the operations phase is shown in Figure 1. The Science Team is composed of all scientific members of all Consortium and Participating Community Surveys; it is the primary exploiter of 4MOST data. The Science Coordination Board represents the 4MOST Surveys and consists of all

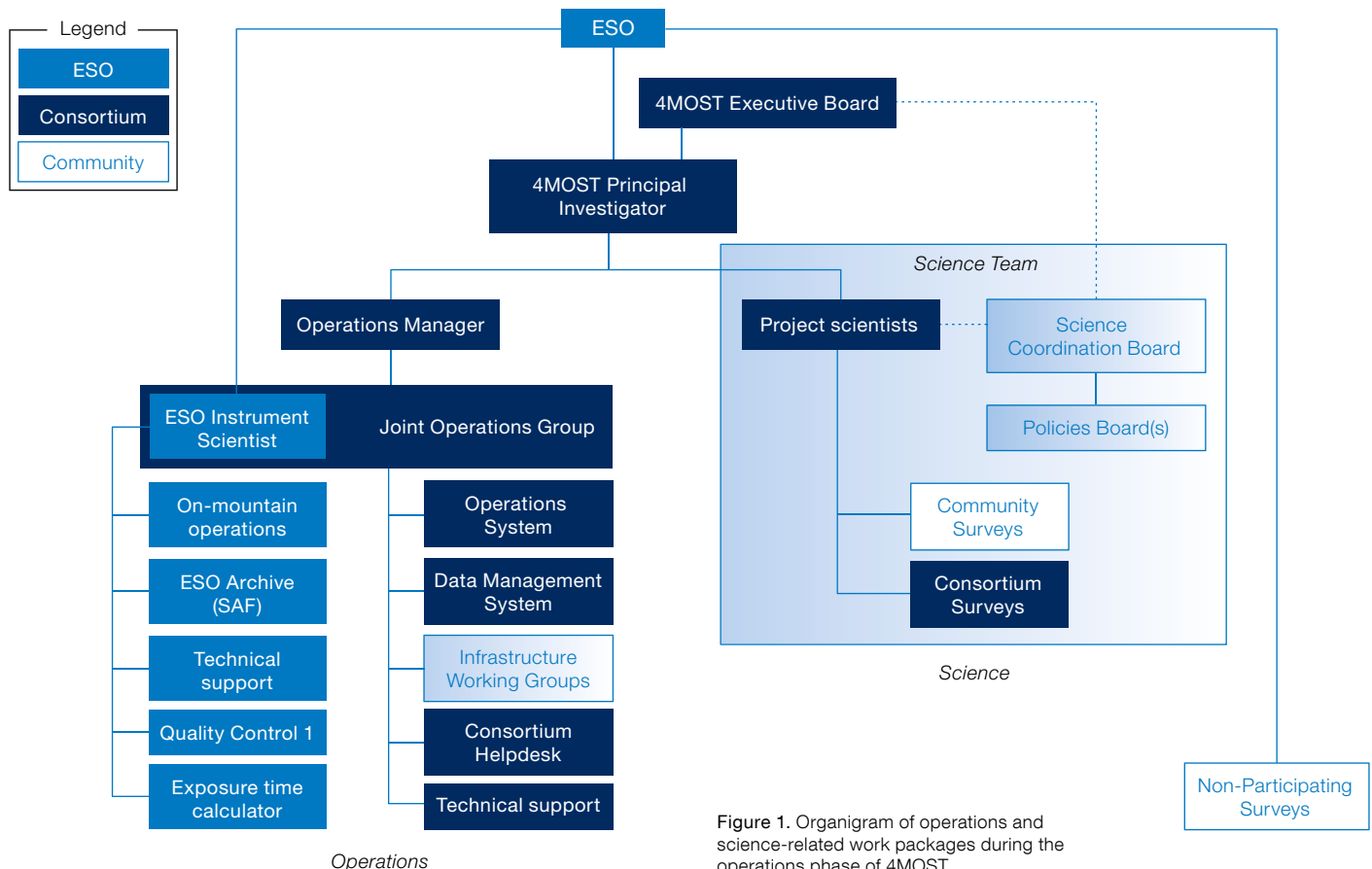

**Figure 1.** Organigram of operations and science-related work packages during the operations phase of 4MOST.

Consortium and Community Survey Principal Investigators (Survey PIs). It coordinates the scientific programmes of the surveys, in the spirit of a single overall 4MOST survey programme. It mediates potential conflicts of interest, and enforces the Science Team Policies. The day-to-day implementation of the Science Team Policies is delegated to the Science Policy Board.

Coordination of the diverse group of stakeholders for 4MOST operations is ensured by the Joint Operations Group. The Joint Operations Group implements the scientific and operational guidance given by ESO, the Survey PIs, and the 4MOST PI. It contains representatives from the various work packages in operation. Its main task is to ensure observations progress with the best possible quality over the entire survey duration.

Infrastructure Working Groups perform tasks common to all surveys. Their members are also members of the Science Team and are delegated to operations. Every Participating Survey is requested to provide resources to the Infrastructure Working Groups. The current working groups are: "Survey Strategy"; "Selection Functions"; "Galactic Pipeline"; "Extragalactic Pipeline"; and "Classification Pipeline". Work in these groups has already started and Community Participating Surveys will be invited to join. This means not only profiting from the work already done, but also providing resources for ongoing development and operations.

### Science Data Flow for Participating Surveys

The data flow through the 4MOST facility (see Figure 2) will follow concepts that are familiar to most astronomers working with either ESO or large survey projects. In a nutshell, the following steps in the data and workflows are foreseen:
1. Preparation of target catalogues with relevant associated data (for example, figures of merit) by surveys.
2. Submission of target catalogues to the Operations System.
3. Merging of catalogues and preparation of OBs by the Operations System.
4. Submission of OBs to ESO.
5. Execution of OBs at the Visible and Infrared Survey Telescope for Astronomy (VISTA) by ESO.
6. Transfer of raw data from the telescope to the Data Management System.
7. Data reduction from raw data to calibrated spectra by the Level 1 pipeline of the Data Management System.
8. Transfer of Level 1 data to advanced pipelines.
9. Data analysis and creation of Level 2 data products by advanced pipelines.
10. Transfer of all data to archives.
11. Science exploitation by surveys and the world-wide community.

### Feedback loops

There are three feedback loops in this workflow:
1. Quality Control 0 will be carried out directly after an OB has been completed at the telescope. Quality Control 0 will only verify that the OB has been executed successfully in a technical sense. Even if some atmospheric or other conditions have not been met in a completely observed OB, it may still yield data for numerous targets where sufficient photons were received to fit the requirements. For targets where the required S/N has not been met, the exposure time still required can be optimised in a later re-observation of the same field.
2. The Data Management System will assess the observational progress per target as part of the Level 1 pipeline. This assessment is based on spectral success criteria provided by each survey. Progress per target will be communicated to the Operations System in order to be taken into account during the preparation of the next round of OBs.
3. The Operations System of the Consortium will provide a progress database allowing surveys and ESO to monitor observational survey progress at least fortnightly, using metrics such as a figure of merit or the number of successfully observed targets. If a survey falls behind expectations, changes to the survey strategy are possible with the approval of the Science Coordination Board and ESO.

### Observing preparations

The Operations System is in charge of preparing the OBs for the 4MOST Participating Surveys. The target catalogues of the Participating Community Surveys will be merged with those of the Consortium, and through an iterative process a joint survey plan will be developed to observe all targets. The Operations System provides tools to estimate the feasibility and likely success of 4MOST observations. Its exposure time calculator delivers the same results as the one ESO provides, but is optimised such that hundreds of thousands of spectra can be treated jointly.

The 4MOST Facility Simulator is able to simulate the operation of 4MOST, including instrument performance, observatory processes, and weather patterns. Both the simulator and the OB builder use the same target catalogues. Additionally, to enable the planning of a common survey strategy, the surveys are required to deliver Spectral Success Criteria and figures of merit. The former are used to determine exposure times for each target. The latter are used to evaluate how well a specific survey has been able to meet its science goals. More details about the tools and procedures to develop a common Survey Strategy Plan are given in the 4MOST survey plan white paper by Guiglion et al. (p. 17).

The Operations System will prepare OBs approximately every three days, taking into account the latest information from the Data Management System on targets that have already been (partially) completed in previous observations. OBs are generated for the next two weeks providing redundancy in case of a connection breakdown between Europe and Paranal. The regular updates of OBs allow for optimised efficiency in observations. The Operations System also hosts a progress database that contains the current observational status for all Participating Surveys (see above).

### Data reduction and analysis

The Data Management System is in charge of developing and running the Level 1 data reduction pipeline for all surveys. This pipeline will remove the instrumental signatures and calibrate the raw data. It will produce all Level 1 data products, including the 1D spectra, their associated variances and bad pixel masks as well as any other associated information. The data reduction pipeline will also generate per-target progress information to be used by the Operations System in its progress monitor and for the preparation of future observations.

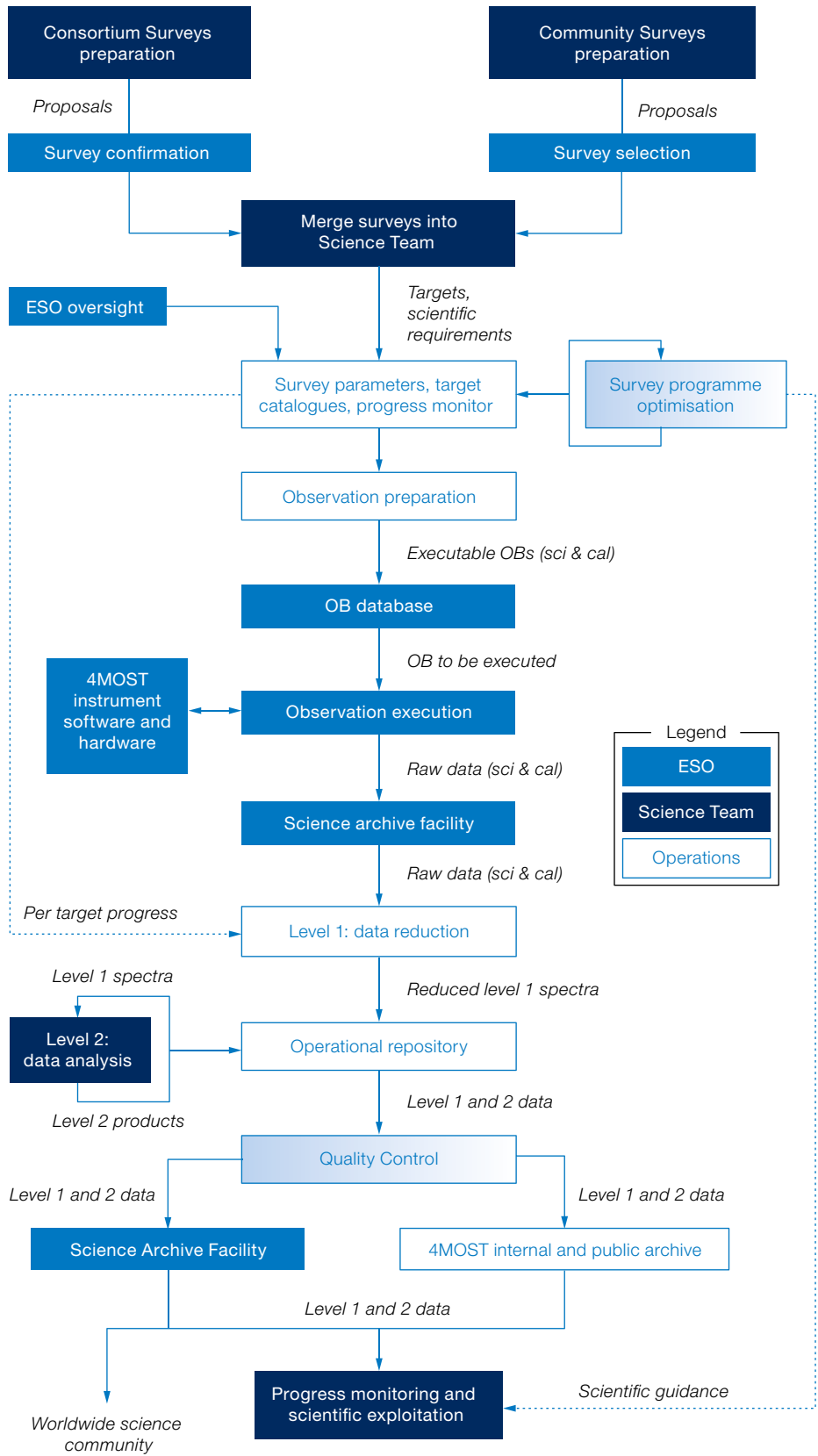

Figure 2. Schematic overview of work and data flows of the 4MOST project during survey operations for Participating Surveys.

The Level 1 data products are primarily used by the advanced pipelines, which produce the advanced Level 2 data products. Currently, there are four advanced pipelines: the classification pipeline; the Galactic pipeline; the extragalactic pipeline; and the selection functions pipeline. These pipelines are developed in Infrastructure Working Groups (see above). The outputs from the advanced pipelines are described in Data eXchange Unit (DXU) documents, which are available on the 4MOST website.

In brief, the classification pipeline will provide a data-driven classification of spectra into stars, galaxies, AGN, outliers and unclassifiable objects. The Galactic pipeline will derive effective temperatures, surface gravities, and element abundances for cool stars, hot stars, and white dwarfs. The extragalactic pipeline will deliver object redshifts, as well as emission line fluxes and stellar population properties. Finally, the selection function pipeline will compute two selection functions — geometric and target. The geometric one can be used to correct for the incompleteness incurred owing to the size and shape of the field of view, tiling, and fibre positioner properties. The target selection function can be used to correct for the effects of finite exposure times and the throughput of 4MOST. Surveys can fold both of these selection functions with the specific target selection procedure per survey, providing an incentive to use a clean and reproducible survey selection function.

The Data Management System will store all data products (Levels 1 and 2)

in a 4MOST Operational Repository. The quality control of data in the repository will be carried out by a working group coordinated by the Quality Control Scientist. The Science Team as a whole will have access to quality controlled data products in three-month intervals as internal releases will be pushed from the repository to the 4MOST Public Archive maintained by the Consortium.

As for all Public Surveys carried out at ESO facilities, the raw data will become public automatically via the ESO Science Archive Facility (SAF). Level 1 data products will be submitted by the Consortium to the SAF and made public by the SAF on a regular basis (nominally yearly). Deliverable Level 2 data products will be submitted to and made public by the SAF on a schedule to be agreed between the 4MOST survey program and ESO. Level 2 data products will naturally lag behind Level 1 spectra by some amount of time that may also depend on the specific survey/data product type. All Level 1 and deliverable Level 2 products will be released not only through the ESO SAF, but also worldwide through the 4MOST Public Archive operated by the Consortium, which in addition may contain matched catalogues from other facilities and added value catalogues with data processed beyond the standard pipelines (additional Level 2 products).

### 4MOST Helpdesk and finding further information

The Consortium-operated 4MOST Helpdesk is tasked with answering questions from all stakeholders in 4MOST (including potential or actual proposers, Participating Survey teams, and users of worldwide data releases). It is reachable by e-mail[1], through an online form accessible from the project webpage[2], and will also serve as a back office for the ESO User Support Department in respect of 4MOST-related questions. The 4MOST Helpdesk is operated by Consortium members and maintains a webpage with frequently asked questions. The webpages of the 4MOST Consortium complement the white papers in this issue of The Messenger with more information. Some key documents will be made available through the webpages, such as the Science Team Policies and details on the planned advanced data products.

### Non-Participating Surveys: a special mode of operation

4MOST has been designed to cover the southern hemisphere in a five-year survey using a parallel mode of observation, enabling surveys that would otherwise not be possible. All Consortium Surveys will be carried out in the parallel mode of observations. At the same time, 4MOST is also a very powerful instrument that can be used in single survey mode if the target density is sufficiently high. In this mode, special thought has to be given to the use of fibres from both Low- and High-Resolution Spectrographs. Proposed Community Surveys wishing to use 4MOST in single survey mode are called Non-Participating Surveys. They will not become members of the Science Team, and will not be bound by the Science Team Policies. Their time will be allocated in named nights or half-nights to enable accurate planning. The Consortium Operations System will deliver the software necessary to produce OBs, which Non-Participating Surveys will run themselves. The Consortium Data Management System will deliver Level 1 data products to Non-Participating Surveys through the Operational Repository. Non-Participating Surveys will not have access to the advanced pipelines developed in the 4MOST Science Team. Non-Participating Surveys will have to produce and upload their own Level 2 products to ESO.

#### Acknowledgements

In addition to the authors of this white paper, the support of many individuals within the 4MOST project and within ESO has been important for the development of the 4MOST operations planning.

#### Links

[1] 4MOST helpdesk: help@4most.eu
[2] 4MOST webpage: www.4MOST.eu

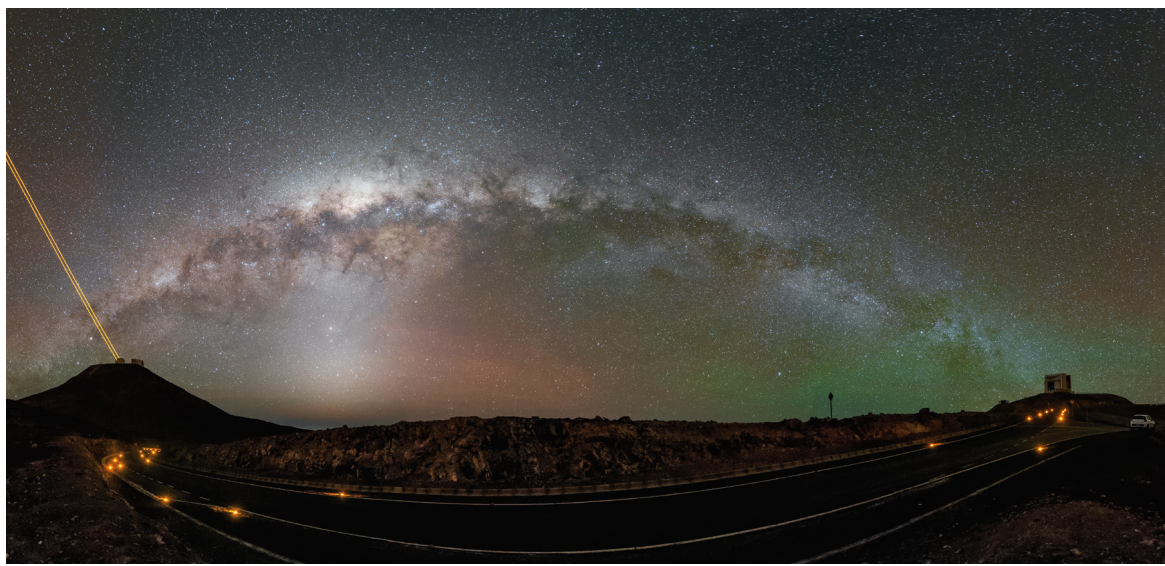

The Milky Way arches over the VLT (clearly deploying its laser guide star capability) and VISTA (on the right). By 2022, VISTA will have transformed into 4MOST with operations beginning towards the end of the year.